COMBINED AND COMPARATIVE TIME-SERIES SPECTRUM ANALYSIS

P.A. Sturrock[1]

[1]Center for Space Science and Astrophysics, Stanford University, Stanford, CA 94305-4060

ABSTRACT

It is often necessary to compare the power spectra of two or more time series. One may, for instance, wish to estimate what the power spectrum of the combined data sets might have been. One might also wish to estimate the significance of a particular peak that shows up in two or more power spectra. Visual comparison can be revealing, but it can also be misleading. This leads one to look for one or more ways of forming statistics, which lend themselves to significance estimation, from two or more power spectra. We here propose two such statistics, one that is most useful for the combined study of two or more similar time series, and another that is more useful for the study of dissimilar time series.

1. INTRODUCTION

Power spectrum analysis is one of the most powerful tools for the study of time series. (See, for instance, Oppenheim, Schafer & Buck 1999.) However, in some studies it would be convenient to combine spectra found from more than one time series and, in other studies, it would be convenient to compare and contrast such spectra. The purpose of this article is to introduce two statistics that are designed to facilitate such analyses.

We introduce in Section 2 a "combined spectrum statistic" that is designed to be used in the former case, where one wishes to combine spectra obtained from two similar but independent time series, and we extend this to higher order in Section 3. We introduce in Section 4 a "joint spectrum statistic" that is intended for use in the latter case, where one wishes to compare spectra obtained from quite different time series, to determine whether they have features in common, and we extend this to higher order in Section 5.

We here adopt the familiar approach of evaluating the probability that a given result may have been obtained by chance on the basis of a "null hypothesis." For the analysis of a single time series, the familiar null hypothesis is that the data are derived from a normally distributed random variable. If the data are normalized to have variance unity, the probability of obtaining a power S or larger, at a given frequency, is given by $e^{-S}$. We consider two ways (summation and multiplication) of combining two or more power spectra. For each procedure, we then derive a function of that combination that has the same exponential distribution. This makes it possible to assign significance levels to these combinations in the same way that one assigns significance levels to a single power spectrum.

## 2. COMBINED SPECTRUM STATISTIC OF SECOND ORDER

For a single time series composed of normally distributed random noise with variance unity, the probability distribution function for the power $S$ (at a specified frequency) is given by

$$P(S)dS = e^{-S}dS. \qquad (2.1)$$

(See, for instance, Scargle 1982.) We see, as noted in the introduction, that the probability of obtaining a power $S$ or more is given by the cumulative distribution function

$$C(S) = \int_S^\infty P(x)dx = e^{-S}. \qquad (2.2)$$

Since S is a function of the frequency $\nu$, $P$ and $C$ also are expressible as functions of $\nu$.

If we are given two time series with power spectra $S_1(\nu)$ and $S_2(\nu)$, we may form the sum of the two powers:

$$Z(\nu) = S_1(\nu) + S_2(\nu). \qquad (2.3)$$

Then the probability distribution function for Z is given by

$$P(Z)dZ = \int_0^\infty \int_0^\infty dx\,dy\, e^{-x-y} \delta(Z - x - y) dZ, \qquad (2.4)$$

which leads to

$$P(Z) = Ze^{-Z}. \qquad (2.5)$$

If we now write

$$\int_Z^\infty P(z)dz = e^{-G}, \qquad (2.6)$$

the "combined spectrum statistic" $G$ is distributed in the same way as the function $C$ defined by equation (2.2). We see from (2.5) and (2.6) that

$$G_2(Z) = Z - \ln(1 + Z), \qquad (2.7)$$

where we now introduce the subscript "2" to clarify that we are here considering a combination of just two power spectra.

## 3. COMBINED SPECTRUM STATISTIC OF HIGHER ORDER

We may extend the procedure of the previous section to the combination of any number of power spectra. If

$$Z = S_1 + S_2 + ... + S_n, \qquad (3.1)$$

we shall prove that the appropriate formula for the corresponding combined spectrum statistic is

$$G_n(Z) = Z - \ln\left(1 + Z + \tfrac{1}{2}Z^2 + ... + \tfrac{1}{(n-1)!}Z^{n-1}\right). \qquad (3.2)$$

We find from (2.6) and (3.1) that the corresponding probability distribution function for $Z$ is

$$P_n(Z) = \tfrac{1}{(n-1)!} Z^{n-1} e^{-Z}. \qquad (3.3)$$

Since

$$P_n(Z)dZ = \int_0^\infty dx_1 ... \int_0^\infty dx_n e^{-x_1 - ... - x_n} \delta(Z - x_1 - ... - x_n)dZ \qquad (3.4)$$

and

$$P_{n+1}(Z)dZ = \int_0^\infty dx_1 ... \int_0^\infty dx_{n+1} e^{-x_1 - ... - x_{n+1}} \delta(Z - x_1 - ... - x_{n+1})dZ, \qquad (3.5)$$

we see that

$$P_{n+1}(Z) = \int_0^\infty dx\, e^{-x} P_n(Z - x). \qquad (3.6)$$

This leads to

$$P_{n+1}(Z) = \frac{1}{n!} Z^n e^{-Z}, \qquad (3.7)$$

which leads in turn to

$$G_{n+1}(Z) = Z - \ln\left(1 + Z + \tfrac{1}{2}Z^2 + ... + \tfrac{1}{n!}Z^n\right). \qquad (3.8)$$

Hence we have justified equation (3.1) by induction.

Figure 1 gives plots of the combined spectrum statistics of orders 2, 3 and 4.

## 4. JOINT SPECTRUM STATISTIC OF SECOND ORDER

We now consider the need to compare spectra from two quite different times series. If one of the time series has very strong peaks and the other has comparatively weak peaks, then simply adding the powers would not be very revealing, since the sum would be dominated by the stronger spectrum. In this situation, it is more useful to form something resembling a "correlation function" by forming the product of the two powers. If we therefore introduce the notation

$$Y = S_1 S_2, \tag{4.1}$$

we see that the probability distribution function for Y is given by

$$P(Y)dY = \int_0^\infty \int_0^\infty du\, dv\, e^{-u-v} \delta(Y - uv) dY. \tag{4.2}$$

As before, it is more useful to introduce the cumulative distribution function,

$$C(Y) = \int_Y^\infty dy P(y). \tag{4.3}$$

If we now define the "joint spectrum statistic" $J$ by

$$C = e^{-J}, \tag{4.4}$$

we see that $J$ is distributed in the same way as the power of a single time series. We find that

$$P(Y) = \int_0^\infty \frac{du}{u} e^{-u-Y/u} \tag{4.5}$$

so that

$$C(Y) = \int_0^\infty du\, e^{-u-Y/u} \tag{4.6}$$

Hence $J$ is given by

$$J = -\ln\left(\int_0^\infty du\, e^{-u-Y/u}\right) \tag{4.7}$$

We find that this is expressible as

$$J = -\ln\left(2Y^{1/2} K_1(2Y^{1/2})\right), \tag{4.8}$$

where $K_1$ is the Bessel function of the second kind.

We see that it would be more convenient to introduce the notation

$$X = (S_1 S_2)^{1/2}, \tag{4.9}$$

so that (4.8) becomes

$$J_2 = -\ln(2X K_1(2X)). \tag{4.10}$$

We here add the subscript "2" to emphasize that statistic is formed from just two power spectra. We find that the leading terms of the asymptotic expression for $J_2$ are as follows:

$$J_2 \to 2X - \tfrac{1}{2}\ln(\pi) - \tfrac{1}{2}\ln(X) \text{ as } X \to \infty. \tag{4.11}$$

Figure 2 shows the joint spectrum statistic $J_2$ as given by (4.10) and by the asymptotic expression (4.11). We see that (4.11) is an excellent approximation to (4.10) for $X > 1$.

## 5. JOINT SPECTRUM STATISTICS OF HIGHER ORDERS

Although it is possible to give a formula for $J_2$ in analytical form, it appears that this is not possible for joint spectrum statistics of higher order. Nevertheless, one can find an iterative relationship between successive orders, and this may prove useful in calculating these functions.

If we write

$$Y = S_1 ... S_n, \tag{5.1}$$

then the probability distribution function for $Y$ is given by

$$P_n(Y)dY = \int_0^\infty dx_1 ... \int_0^\infty dx_n e^{-x_1-...-x_n} \delta(Y - x_1...x_n) dY. \tag{5.2}$$

By examining the corresponding formula for $P_{n+1}(Y)$, we find that

$$P_{n+1}(Y) = \int_0^\infty dx\, e^{-x} P_n(Y/x). \tag{5.3}$$

In terms of the cumulative distribution functions,

$$C_n(Y) = \int_Y^\infty dy\, P_n(y), \tag{5.4}$$

the iterative relation (5.3) becomes

$$C_{n+1}(Y) = \int_0^\infty dx\, e^{-x} C_n(Y/x). \tag{5.5}$$

Probably the best way to use this relation is to find a simple analytical approximation to $C_2$, and use this to estimate $C_3$, etc. Once the cumulative distribution function has been found in this way, one may find the corresponding values of the joint spectrum statistic by using equation (4.4). One would then find it convenient to express the statistic in terms of X, rather than Y where

$$X = Y^{1/n}. \tag{5.6}$$

The advantage of this transformation is that the range of $X$ is then comparable to the range of $J$, so that tables and graphs become more manageable. We find that the asymptotic form of each function $J_n$ has, as its leading term, a multiple of the variable $X$. For $X > 5$, $J_3$ and $J_4$ are given approximately as follows

$$J_3 \to 3X - 3.0, J_4 \to 4X - 4.5, \quad as \quad X \to \infty. \quad (5.7)$$

One may also calculate the joint distribution functions by a Monte Carlo process. If $R$ is a random variable between $0$ and $1$ and $S = -\ln(R)$, then $S$ is distributed as in (2.1). By generating many sets of three or four $S$-values, we have calculated many values of the statistics $J_3$ and $J_4$, and then formed the histograms of these tabulations. Figure 3 shows $J_3$ and $J_4$ as calculated in this way, together with $J_2$ as calculated from (4.10).

## 6. DISCUSSION

We have introduced each statistic as a combination of the powers $S_1(\nu)$ and $S_2(\nu)$ computed for the same frequency from two different data sets. However, they can also be used to combine powers formed from different but related frequencies, from the same or different power spectra. For instance, if we are studying the influence of rotation upon solar variables, we may expect to find a peak in the spectrum at the rotation frequency $\nu_R$ (if it is well defined) and also at harmonics of this frequency. Hence we could choose to form a statistic by combining $S(\nu)$ and $S(2\nu)$, for instance.

Bai (2002) has recently reviewed evidence that solar flares tend to exhibit periodicities with periods that are integer (2, 3, 4, …) multiples of a "fundamental period" that is approximately 25.5 days. The significance of this result could be assessed by means of a statistic formed from $S(\nu/2)$, $S(\nu/3)$, $S(\nu/4)$, etc. Wolff (2002) has recently claimed to find somewhat similar patterns in power spectra formed from measurements of the solar radio flux. The statistics we have here introduced should be helpful in the further evaluation of such claims.

We have used these statistics in the analysis of Homestake (Cleveland et al. 1998) and GALLEX-GNO (Anselmann et al. 1993, 1995; Hampel et al. 1996, 1997; Altmann et al. 2000) solar neutrino data to search for rotational periodicities that are common to both data sets, and to seek and evaluate evidence of r-mode oscillations (see, for instance, Saio 1982) in these data. These analyses will be published separately at a later date.

This research was supported by NSF grant ATM-0097128.

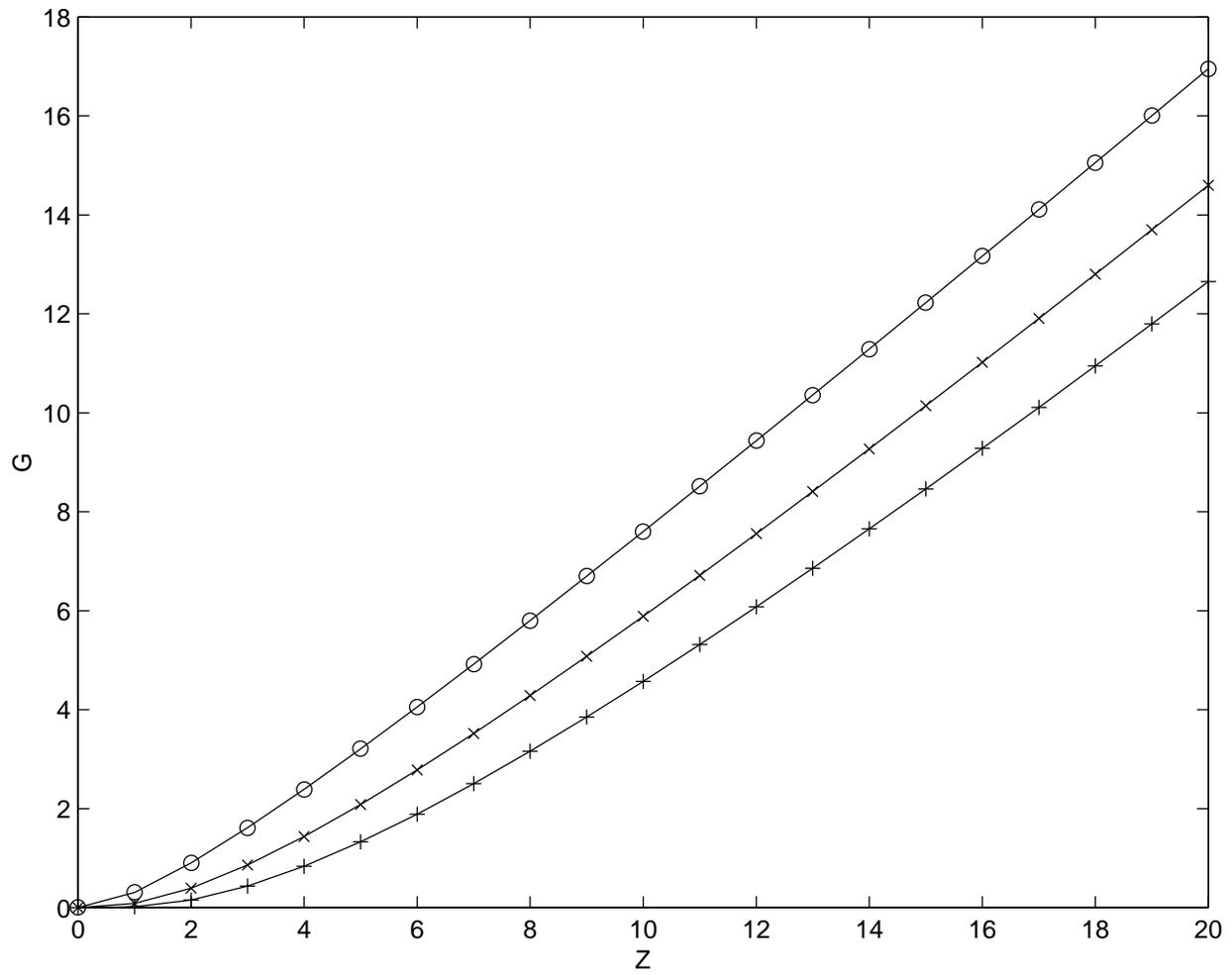

Figure 1. Combined spectrum statistics of second, third and fourth order: G2 ('o'), G3 ('x'), and G4 ('+').

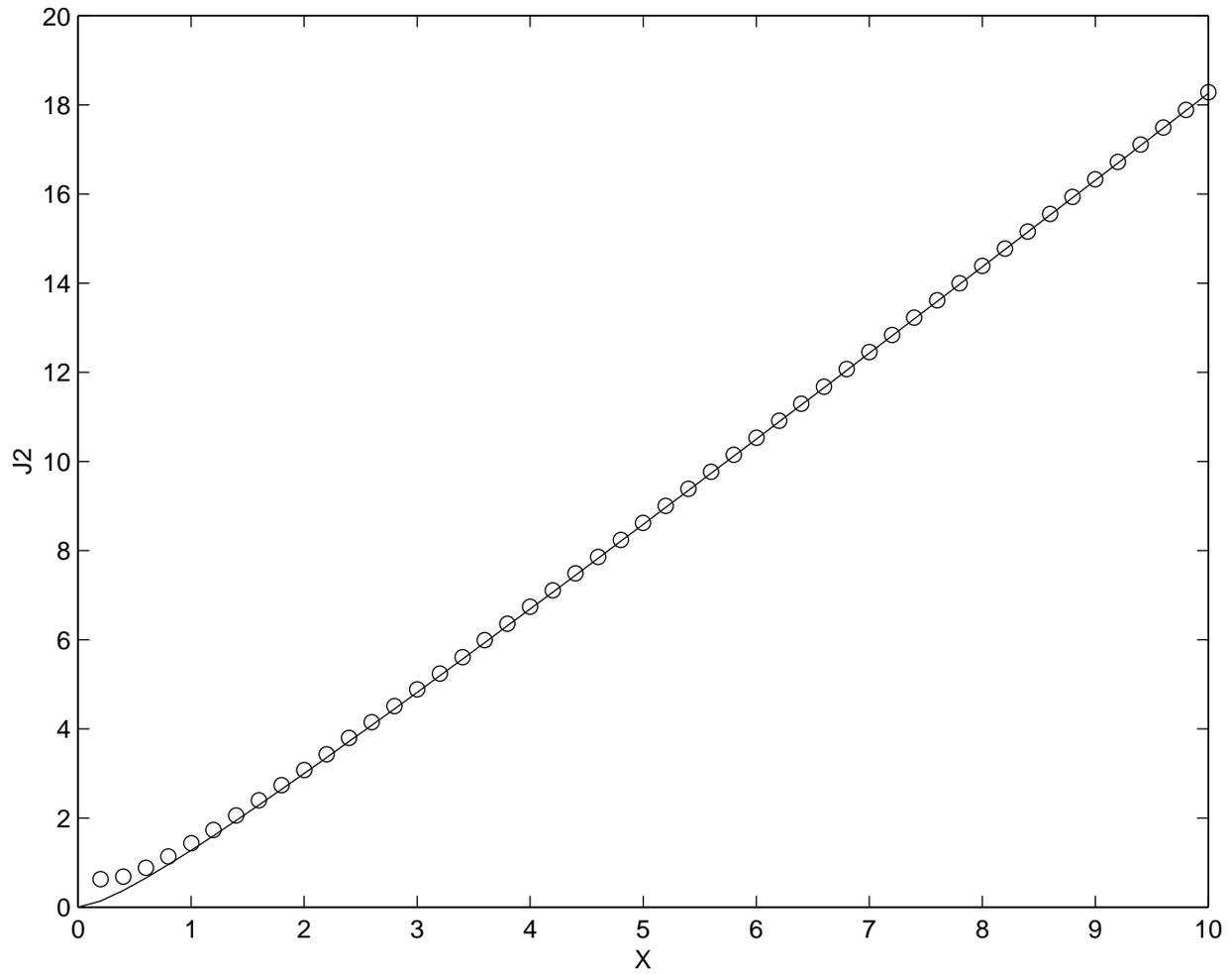

Figure 2. Joint spectrum statistic of second order as given by the analytical expression (4.10) (solid line) and as given by the asymptotic expression (4.11) (circles).

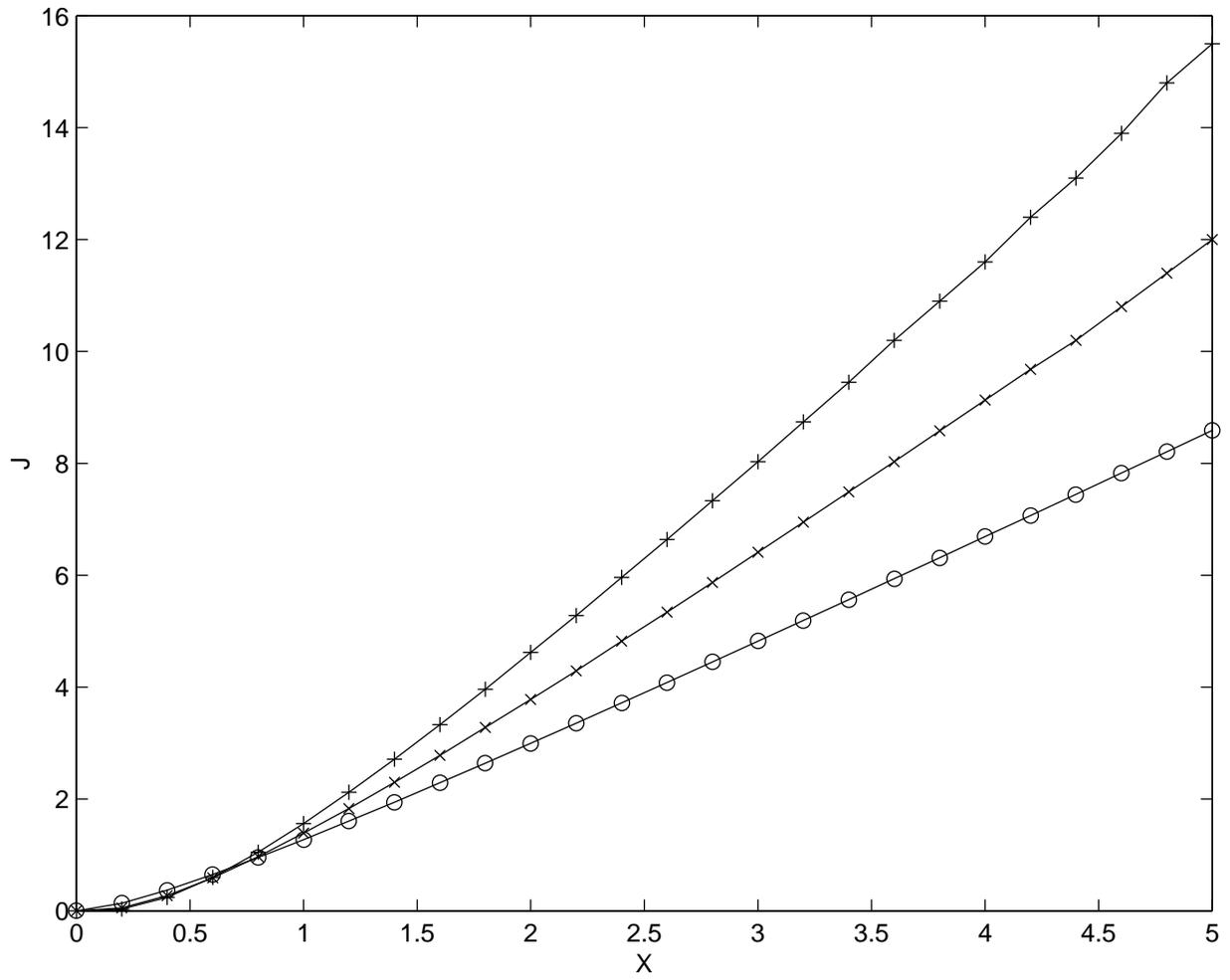

Figure 3. Joint spectrum statistics of orders 2, 3 and 4: J2 ('o'), J3 ('x'), J4 ('+').